\documentclass[amsmath,amssymb,aps]{revtex4}

\usepackage{graphicx}% Include figure files

\begin{document}
\title{On the Discovery of the GZK Cut-off.}

\author{Tadeusz Wibig}
 \email{wibig@zpk.u.lodz.pl}
\affiliation{%
Physics Dept., University of Lodz;\\
Cosmic Ray Lab., Soltan Institute for Nuclear Studies, Uniwersytecka 5,
90-950 Lodz, Poland.}

\date{\today}
\begin{abstract}
The recent claim of the '5 sigma' observation of the Greisen and Zatzepin and Kuzmin
cut-off by the HiRes group based on their nine years data is a 
significant step toward the eventual solution of the one of the 
most intriguing questions which has been present in physics for 
more than forty years. However the word 'significance' is used 
in the mentioned paper in the sense which is not quite 	
obvious. 
In the present paper
we persuade that this claim is a little premature.

\end{abstract}

\maketitle

\section{Introduction}

According to the usual practise the `5 sigma confidence level 
(the chance probability of occurrence of about 1:10000000) is the 
level of discovery in physics. An observation of something which by chance 
can appear so rarely gives as all rights to believe that 
there is some cause, yet unknown, but worth to be studied further.
The discussion of the case of recent GZK cut-off discovery is the subject the present paper.  

After almost 100 years of research, the origin of cosmic rays is still an 
open question. The CR energy spectrum exhibits little structure and is approximated by broken
power laws. The first break, called the ``knee'', appears at the energy of 
$E \approx 4 \times 10^{15} $eV , the flux of particles steepens 
from a power law index of about 2.7 to one of index 3.0.
The bulk of the CRs up to at least that energy is believed to originate 
within the
Galaxy. The spectrum continues with a
further steepening to $\sim$3.3 at $E \approx 4 \times 10^{17}$eV, 
sometimes called the ``second knee''. There are
some indications, see, e.g., Ref.~\cite{hor}, that 
the mass composition changes from light at the
knee to heavy dominated by iron nuclei at the second knee. This is expected
if acceleration and propagation is due to magnetic 
fields only and 
depends on particle rigidity.

The second knee could be related to the transition from the Galactic 
cosmic ray flux to the extragalactic one around the ``dip'' structure 
at $E \approx 5 \times 10^{18}$~eV. Some argue that this switch takes place
before \cite{my}, and some that after \cite{bere} the dip, but anyhow, the spectrum then flattens 
again to a power law with
an index of $\sim 2.8$ forming the so-called ``ankle''.

The possibility of another change of CR composition around the ankle is 
a subject of extensive experimental studies. 
If there is a proton dominated extragalactic cosmic ray flux
above the ankle, then, according to the works of 
Greisen \cite{Greisen-1966-PRL-16-748}, and Zatsepin and
Kuzmin \cite{Zatsepin-1966-JETPL-4-78} (GZK) published in 1966 just after 
the Cosmic Microwave  Background (CMB) had been discovered, 
the suppression in cosmic-ray flux beyond certain energy 
is inevitable.
The mechanism of this cut-off is that protons travelling intergalactic 
distances would interact with the CMB photons
losing energy producing the $\Delta^+$ resonance. A energy threshold for this process
was predicted in 1966 at $\sim{6}\times{10}^{19}$~eV. 

The observation of sharp 
cut-off of the CR flux around this energy 
gives the strong evidence of the proton dominant composition 
of the Ultra High-Energy Cosmic Ray (UHECR).

On the other side, the observation of the absence of the cut-off 
is no less meaningful. 
It means that the particles there, because there should be no protons, ought
to be heavy nuclei or something even much more exotic. The heavy nuclei doubtlessly
disprove any top-down mechanisms of UHECR particle creation.

Of course there is still a possibility that all extragalactic UHECR 
(including protons above the GZK threshold) are coming from distances
less than $\sim 50$~Mpc which is very close in cosmological scales.

The HiRes experiment group claim recently 
that the idea of Greisen and Zatzepin and Kuzmin was confirmed with 
the statistical
significance of 5 standard deviations.

From the experimental point of view the determination of particle energy
is based on 
well known for more than sixty years fact that the particle entering the
atmosphere initiates the cascade of secondary particles
created in the chain of subsequent interactions. This cascade, called Extensive Air Shower (EAS)
consists
a huge number of charged particles (mostly electrons and positrons) 
which lose energy exciting the atmospheric particles. They in turn 
could then emit light, mainly in the UV region, and this light could be, 
in principle, registered. 
Registration of this scintillation light flashes is the one way of 
counting the number of charged particles in EAS. Another one is to count the
particles reaching the ground with the number of detectors spread over the wide area.
The number of particles, shower size, is related of course to the total primary energy
of the UHECR particle. The problem how to transform the EAS size
in each individual case to the primary particle energy, or the whole measured size
spectrum to primary energy spectrum is the subject of extensive simulation
studies. It is believed that our knowledge allows experimentalists to perform such
transformation with some reasonable accuracy (of order of about 20\%).

The HiRes project is using the fluorescent technique. It has been described, e.g., in
\cite{AbuZayyad-1999-ICRC-26-5-349,Boyer-2002-NIMA-482-457}.  The
experiment consists of two detector stations (HiRes-I and HiRes-II)
located on the  Dugway Desert  in Utah, US, 12.6 km apart.
Each station is assembled from telescope modules (22 at HiRes-I and 42
at HiRes-II) pointing at different parts of the sky, covering nearly
$360^{\circ}$ in azimuth, and $3^{\circ}$--$17^{\circ}$ (HiRes-I), and
$3^{\circ}$--$31^{\circ}$ (Hires-II) in elevation.  Each telescope
module collects light from air showers using a
spherical mirror of about 4 ~m$^2$ area. The camera for each telescope
is a cluster of
photomultipliers of the field of view of a $1^{\circ}$ diameter cone on the sky
\cite{Sokolsky-2007-ICRC-30-1262}.

The data of 
HiRes consists of three sets: two of them are the 
monocular data collected by HiRes-I and HiRes-II detector 
stations, and the third is the set of events registered 
simultaneously by both stations. The smaller statistics of the 
last one is related to the high energy threshold for 
the showers to be seen by both stations as well as the limited 
geometry and thus effective collection area. However the stereo 
observation makes the energy estimation much more accurate and 
gives the possibility to check the mono-eye reconstruction 
procedures for systematic biases at least in the limited sample 
of stereo events. 

The Ref.~\cite{Abbasi:2007sv} conclusions are based only the 
monocular HiRes data from both stations (the showers seen by  
both were excluded from the HiRes-I sample to preserved the  
statistical independence of both station measurements). We would like 
to look as well to the stereo data which can be found, 
e.g., in Ref.~\cite{Sokolsky:2007hf}.

We don't want also to discuss here also all the details concerning 
energy determination procedures, sources of uncertainties, also 
these systematic, as well as the complicated question of the 
apertures estimation. These question were subjects of the 
extensive analysis made by the HiRes team for years. The 
published statements make the procedures as trustful as one can 
get, taking into account the inevitable knowledge of the very 
high interaction mechanism, still not known very precisely  
fluorescence yield, the status of the atmosphere in every 
particular case, possible uncertainties, oversimplifications of 
the simulation and reconstruction programs, to name only few 
possible sources of experimental difficulties, not to mention 
the problems of the hardware nature.

\section{The probability \label{proba}}

The paper \cite{Abbasi:2007sv} can be used as an example of 
misunderstanding, or rather, as an illustration of the general 
problem with the concept of {\em probability}. 

There are at least two (main) interpretations of the 
probability itself. One called {\em classical} is well known and obvious,
but it should be 
remembered that it is common not for very long time. 
Approximately, as classical as 'classical' is the special theory of relativity
(since the times of Pearson), or even as old as quantum physics (since 
Neyman or Fisher milestone papers). 
Sir Francis Galton the pioneer of the statistical treatment of the data, 
was knighted less then hundred years ago, in the year 1909.
The frequentalist meaning of the {\em probability} is given 
already at schools. In the common form, as a limit of the fraction 
of successes in the infinitely long sequence of {\it 
identical} trials. (The 'identity' should be understand as the
requirement  
that the 'probability' in question remains constant during the 
sequence of trials. This {\it circulus in definiendo} is one of nightmares 
of the frequentalists.)

Another way of thinking about the probability comes from Thomas  
Bayes and it is more than a hundred years older, however, in its 
modern form as "Bayesianism" has been used since about 1950.
The Bayesian point of view stands for the probability as a rate  
of rational bet, a degree of belief

The Bayesian treatment of the probability makes it
closer to the common sense. People don't have to have the big
(infinite) sample of results of the repeated experiment 
(in exactly the same conditions etc.) to say something about, 
e.g., the Higgs boson mass, or the appearance of the sun
tomorrow morning. 

The Bayesian definition of probability is 'by definition' 
subjective. The probability itself doesn't exist, in a sense 
\cite{agostini}. The most important for the discussed problem is the 
fact that the Bayesian probability of the event is defined in a 
certain moment in time. When the time passed, the value of the
probability (the rational bet) may change caused by the 
increase or decrease (we'll come back to this last intriguing
possibility in Sec.~\ref{zapo}): a change, in general, of the 
information about the subject one has gathered in the meantime.

This situation is common in physics. The progress of our 
understanding of the Universe is expected to be related to 
the new experiences, 
observations, measurements. All of them change a background 
which makes the base of the estimation of our rational bets on 
reality (to be made, e.g., for further experiment outputs). 

The Bayesian analysis is based on his famous theorem, which can be expressed as:
$
P(H|E) \sim P(H) \; P(E|H)
$,
where the proportionality constant is determined by the 
normalisation of the $\sum _i P(H_i)P(E|H_i)$. 
The first factor $P(H)$ is the probability of the hypothesis $H$ 
to be true prior to the experiment output $E$
is known. The $P(E|H)$ is the likelihood of the output $E$ if the 
hypothesis $H$ is true. The left side of the
equation is the improved new probability of the hypothesis $H$ to 
be true if we know the result $E$.

The question of the existence of the GZK cut-off can be answered in terms of
probability. The proposition "there is a sharp cut-off of the very high energy
cosmic ray spectrum" and the opposite "no such cut-off exists". These are 
statements about the reality and of course only one of them can be true.
To judge this in a scientific ('classical') way
one has to test the GZK cut-off hypothesis statistically. 
The standard, Fisher or frequentalists, answer to the test question can be
only that at given {\em confidence level} there are no observational 
constrains to the GZK hypothesis,
or the hypothesis should be, according to the performed observations, 
rejected on this {\em confidence level}.
The Bayesian answer can be that there {\em is} a given probability,
estimated according to all the knowledge we have, that the 
GZK hypothesis is true (or false, if one wish).

We will discuss this difference. 

\section{The prior \label{sprior}}

The question if the cosmic ray energy power-law spectrum extends
continuously with more or less constant index stands before the
famous Greisen, and Zatsepin and Kuzmin papers predicting a sharp end of this
spectrum around few (6) times 10$^{19}$~eV appeared. The GZK cut-off as a result
of interaction of Ultra High-Energy Cosmic Ray protons with 
Cosmic Microwave Background (CMB) photons,
couldn't be proposed before the CMB radiation itself was discovered in 1965. 
But even then UHECR were intriguing due to the fact that they,
according to the great magnetic rigidity, couldn't be confined 
within the Galaxy or in other known Galactic object.

 \begin{figure}[th]
\centerline{
\includegraphics[width=7cm]{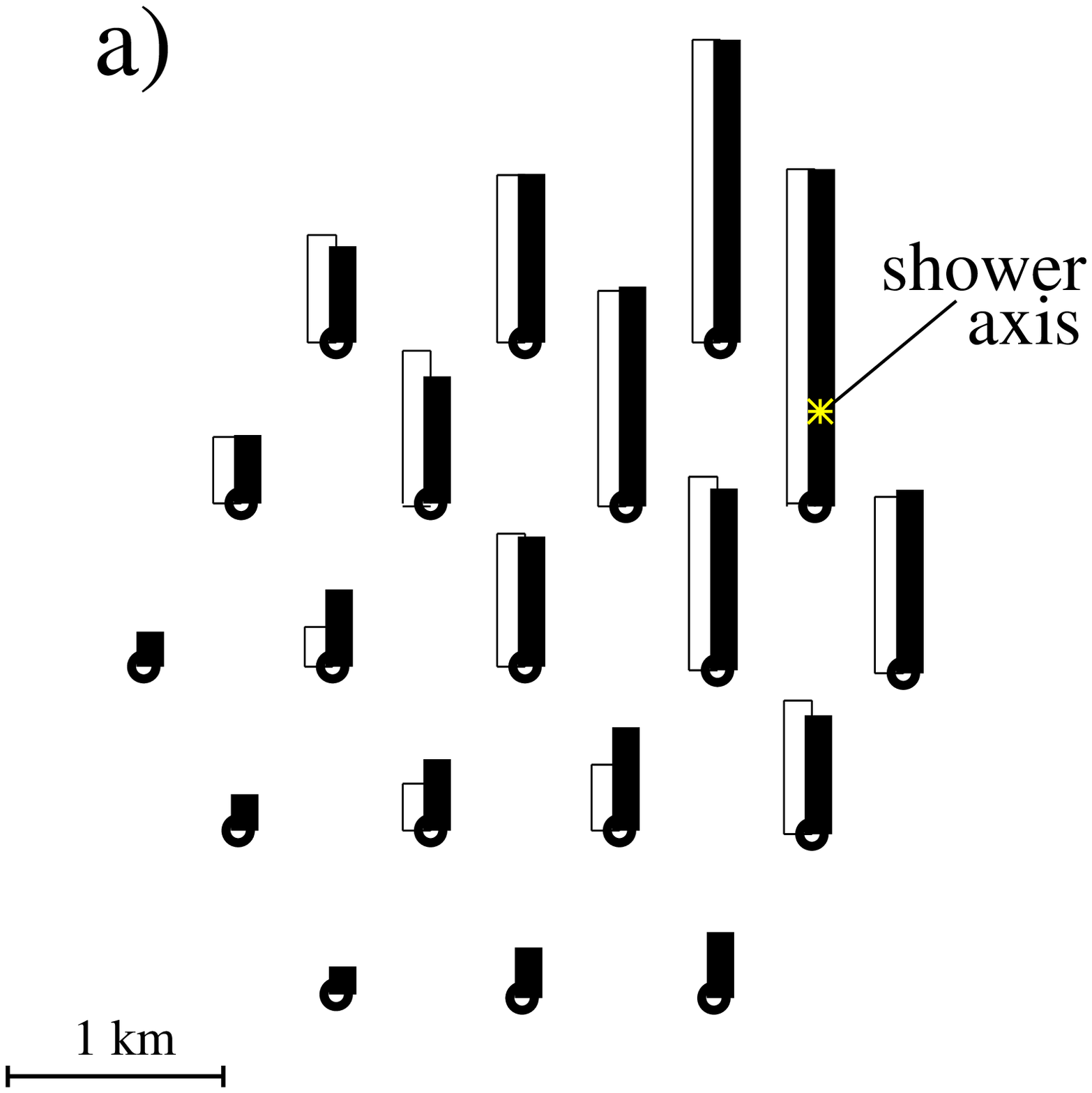}
\includegraphics[width=7cm]{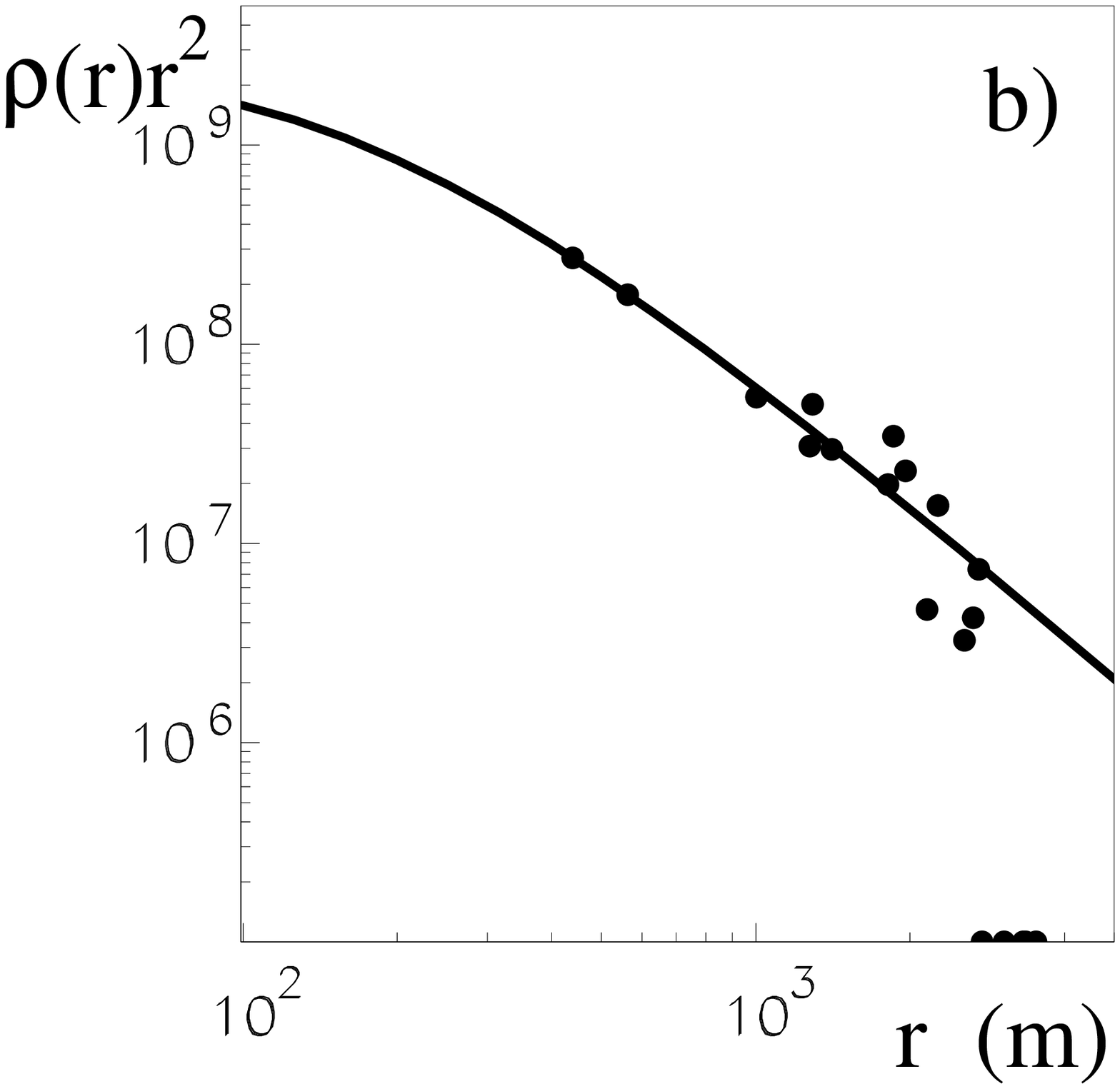}
}
\caption{The density map a) of the Volcano Ranch super-GZK event No.~2-4834 of energy estimated as $10^{20}$~eV. Positions of detectors are shown by circles and the blank bars beside represent the respective registered density (its logarithm)
while the filled bars show densities obtained with the 'best fit' lateral shower particles distribution (the NKG formula) shown as a function of the distance to the shower axis in b). The adjusted position of the axis is shown in a). 
\label{fig1}}
\end{figure}

The sharpness of the CR energy spectrum require EAS arrays 
of bigger and bigger areas to register the highest energy particles.
The first really great one dedicated for the UHECR domain was the Volcano Ranch array.
It has been running since 1960.
It consists of 19 3.3 m$^2$ scintillator counters distributed aver about 10 km$^2$
in Dugway. This experiment, relatively simple and small in comparison with 
contemporary projects, registered in February 1962 the event No.~2-4834 \cite{volcano1}.
shown in Fig.~\ref{fig1}a. 11 out of 19 detectors registered particles. The 
highest signal was estimated to be equivalent of about 1400 minimum ionisation 
particles per m$^2$. 
The shower particle lateral distribution 
was found using the form known as NKG-function. Its integration gave the total number of 
charged particles in the shower equal to $5 \times 10^{10}$. There registered 
densities (their logarithms) in comparison with the values  
of the fit are shown in Fig.~\ref{fig1}a as vertical bars along 
each detector positions.
It is seen in Fig.~\ref{fig1}b that the found particle distribution describes the 
points very well. The tests with another distributions used in different experiments
doesn't make it any better and the total number of particles doesn't change more than 
20\%. As an ultimate test we can try to eliminate from the fitting procedure the
stronger signal detector arguing that it can be made by some internal cascading, thus
not representative to the shower particles. The usual NKG function fit however doesn't
change the result.  If one 
release all parameters in NKG-like function (both indexes and radius scale parameter)
the 'best fit' can produce eventually the size of the shower significantly different, 
(N$_{\rm e} = 7 \times 10^{9}$ instead of $5 \times 10^{10}$) with a slight displacement of the
estimated shower core position.
But the shape of this lateral distribution is so different from conventional wisdom 
and it couldn't be
taken as real better shower description. The usual NKG function is well established
experimentally and it is close to the shower particle distribution measured in different
experiments.

\begin{figure}[th]
\centerline{
\includegraphics[width=7cm]{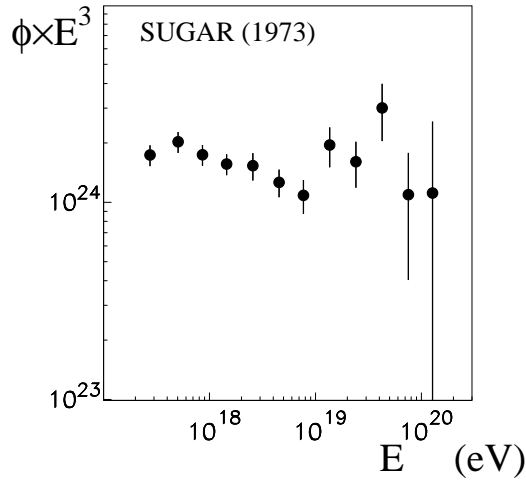}}
\caption{SUGAR 1973 spectrum from \cite{syd1}. 
\label{syd}}
\end{figure}

All this was reminded to cite here explicit the statement given by Linsley in Ref.~\cite{volcano2}:
{\bf "The first observation of the spectrum above $10^{19}$~eV, at Volcano Ranch, showed 
that the spectrum extends to $10^{20}$~eV without a sign of any cut-off."}

Also before the CMB discovery the really very big array was constructed on the Southern hemisphere. 
The Sydney University Giant Airshower Recorder (SUGAR) 
consists of more than 50 stations, each containing two 6 m$^2$ scintillator detectors 
buried underground spread over the surface of about 100 km$^2$ area. 
Detectors of SUGAR were able to register
only EAS muons of energies greater than about 1 GeV. 
The spectrum obtained this way published
in 1973 in Refs.~\cite{syd1,syd86} is shown in Fig.~\ref{syd} .

The paper \cite{syd1} was concluded with the statement: {\bf "It appears likely that
the primary energy spectrum extends beyond $10^{20}$~eV with no significant features...".}

The interest of UHECR increased in the meantime while the CMB was discovered and famous 
papers announcing the GZK cut-off has been published. 

\begin{figure}[bh]
\centerline{
\includegraphics[width=6cm]{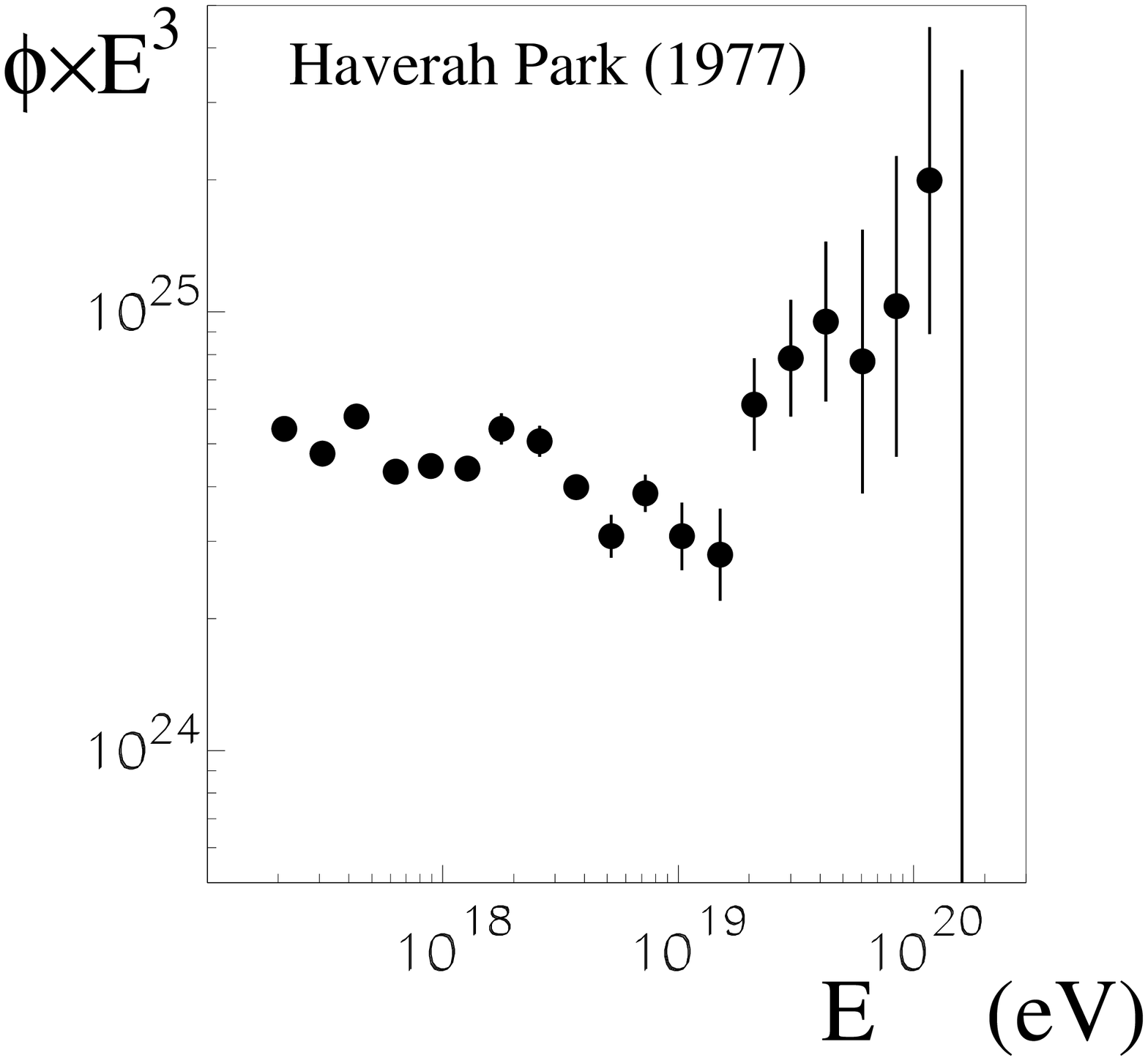}
\includegraphics[width=6cm]{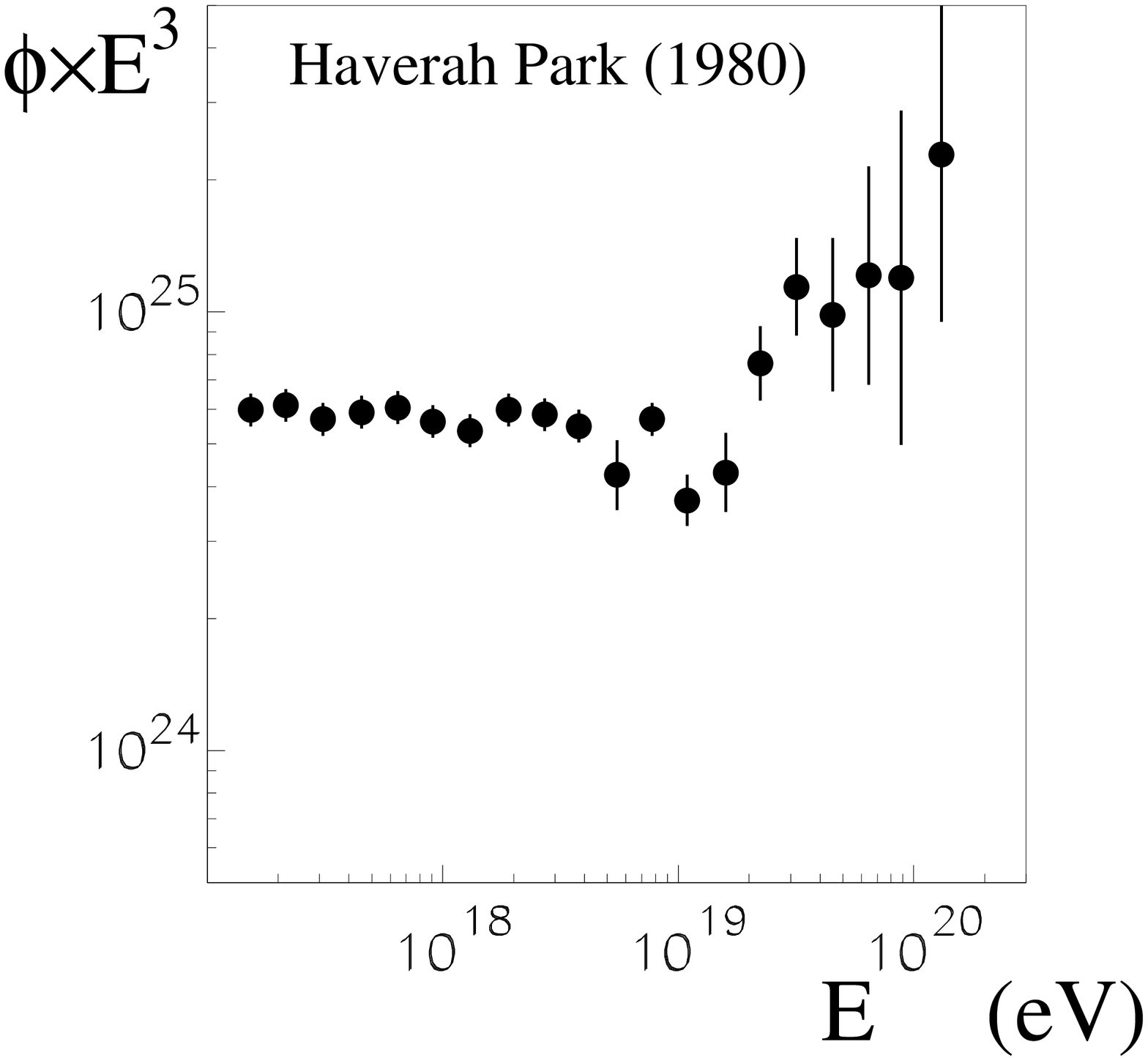}
\includegraphics[width=6cm]{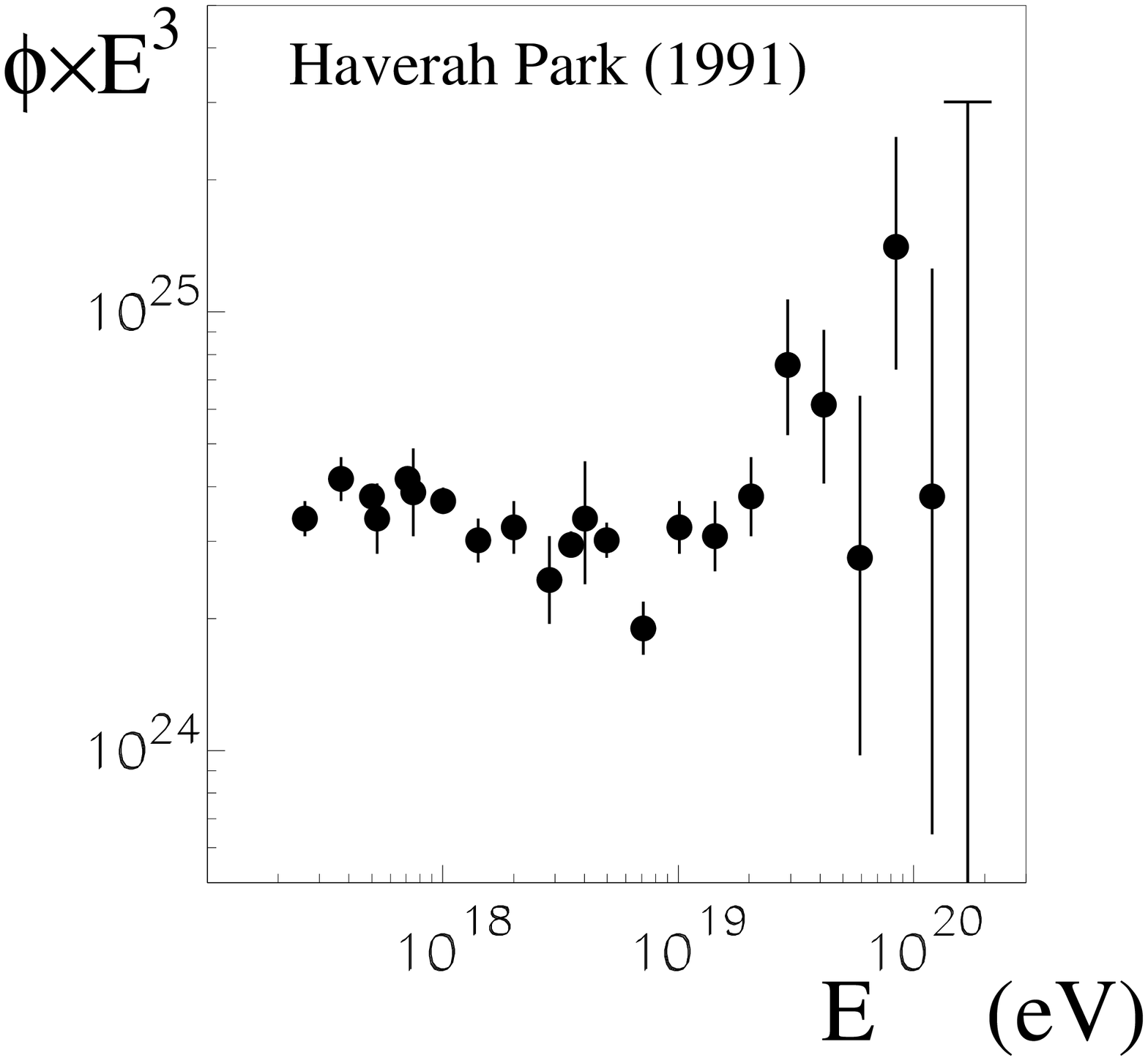}
}
\caption{Haverah Park spectra from \cite{hp77,hp80,hp91}, a), b) and c) respectively. 
\label{hp789}}
\end{figure}

The Northern sky was then monitored by the EAS array built in Haverah Park near Leeds, UK. It reach the size of the 
Volcano Ranch at about 1968. Different types of detectors (water \v{C}erenkov tanks) but also special detectors for muon shower component were installed there.
About 30 years ago the results on the UHECR energy spectra were announced.
We would like to show here three spectra published by the Haverah Park team from the first
published in the end of '70 Refs.~\cite{hp77,hp80} to final 1991 year spectrum \cite{hp91}.
A kind of evolution is seen. It will be discussed later on.

\begin{figure}[bh]
\centerline{
\includegraphics[width=7cm]{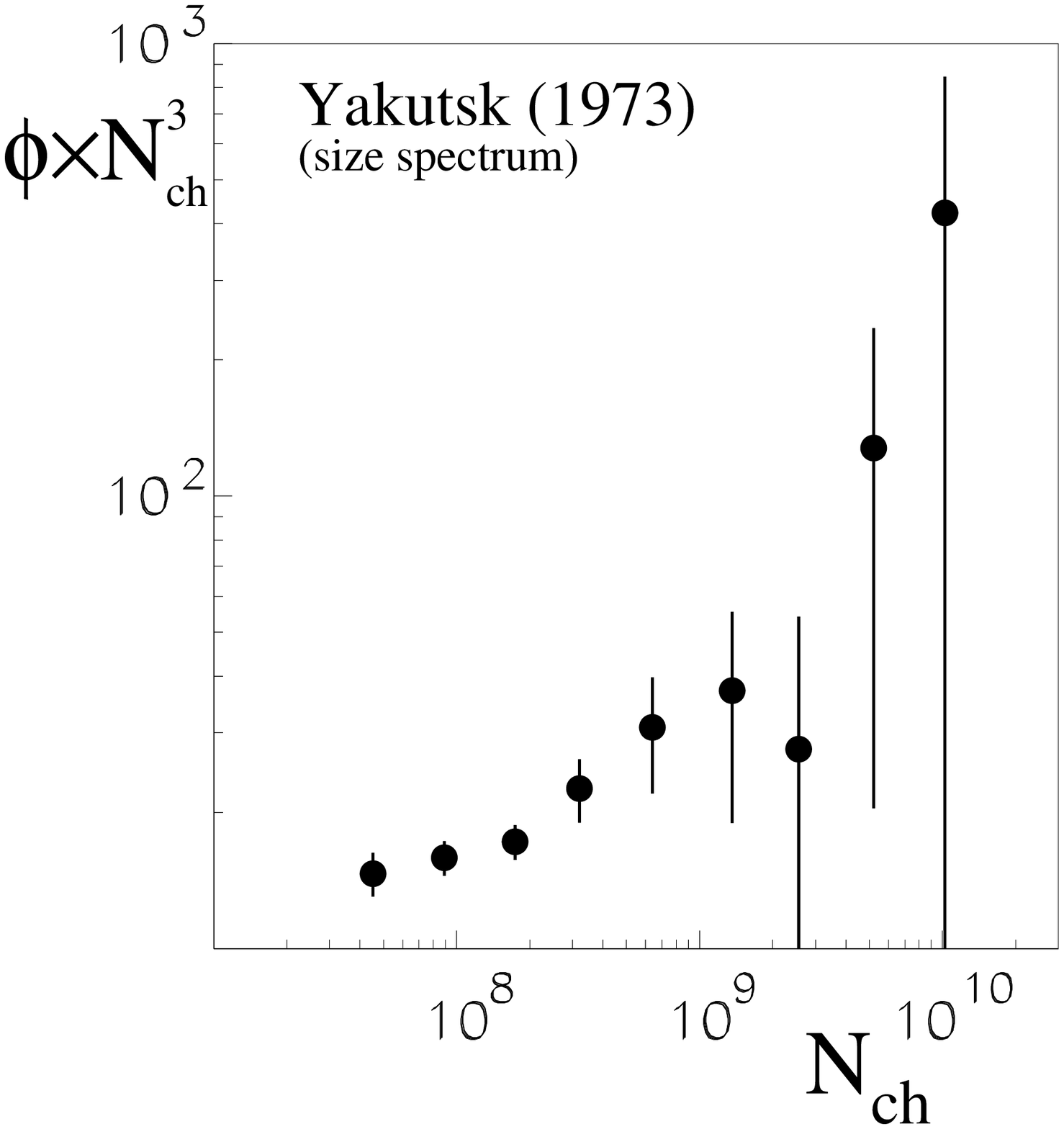}
\includegraphics[width=7cm]{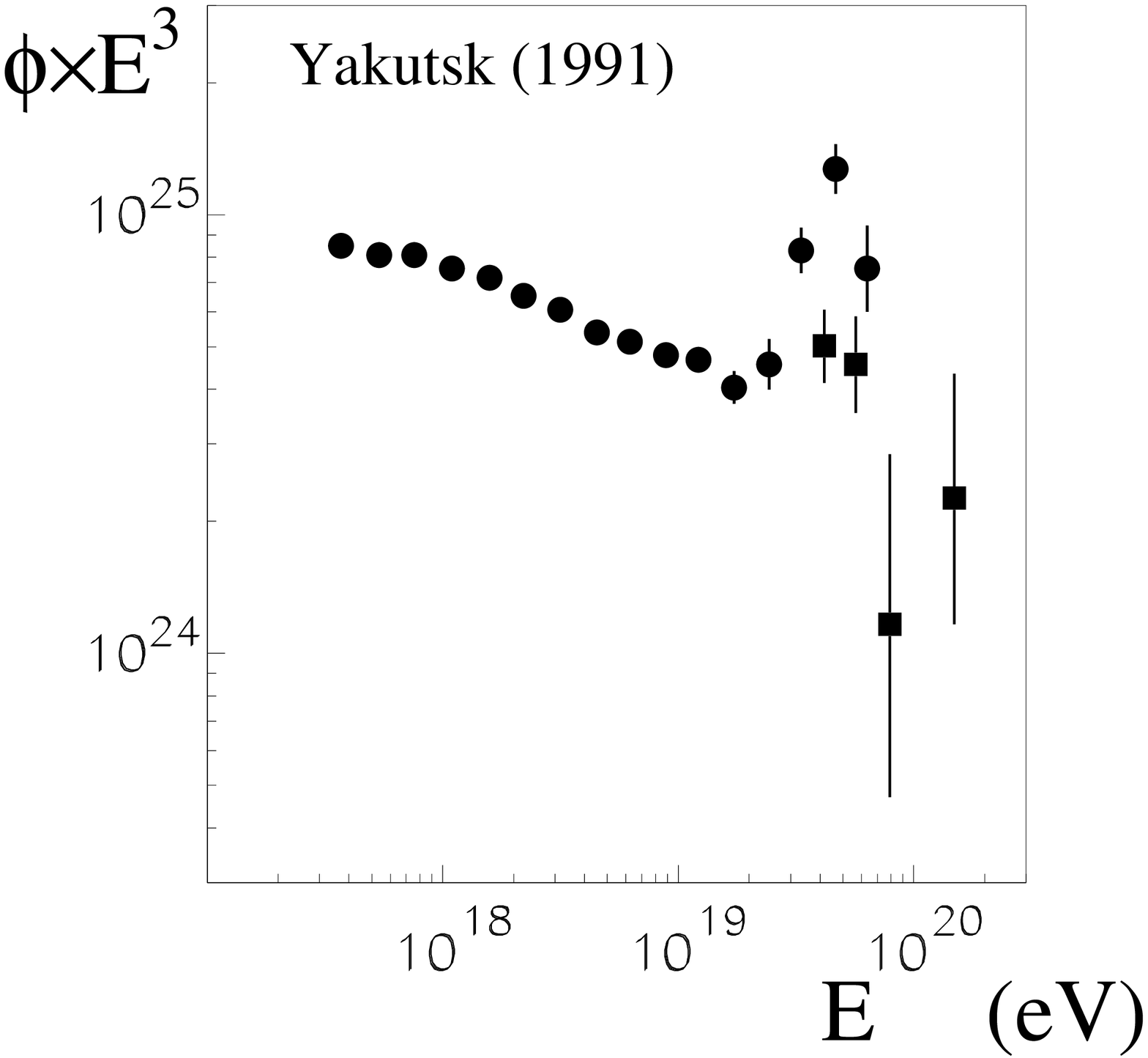}
}
\caption{Yakutsk size spectrum from 
\cite{yak73} (left) and the recent energy spectrum \cite{yakrecent} (right). 
\label{yak}}
\end{figure}

In the beginning of '70 also in USSR the Yakutsk array has started collecting data (and it is still in operation). The first data concerning the size spectra was published in Ref.~\cite{yak73}. It is shown in the Fig.~\ref{yak} together with the recent Yakutsk result published in Ref.~\cite{yakrecent} in 2003.

\begin{figure}[th]
\centerline{
\includegraphics[width=7cm]{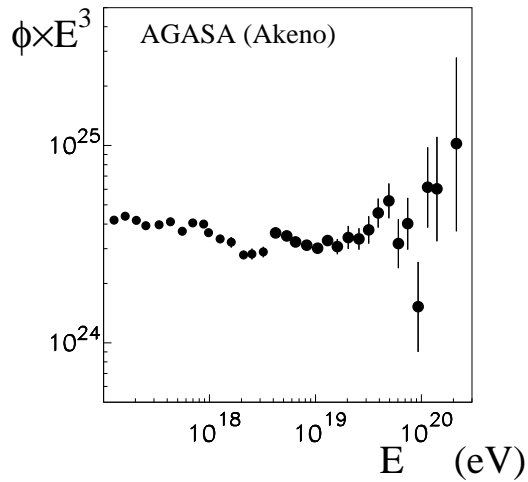}}
\caption{Akeno and AGASA experiment spectra. 
\label{ag}}
\end{figure}

The most controversial (at present) spectrum from the big experiment AGASA \cite{agasas}
is shown in Fig.~\ref{ag}. There are about a dozen of events (the last recorded in 2002) exceeding the GZK limit. 

Clear signal 
presented in the figure is the extreme in the sense that it evaluated with time (mostly due to adjustments of the energy estimation procedures) being less pronounced, but the AGASA result always contradicts the GZK cut-off, more or less strongly. It is worth to mention that in 1993, the AGASA array recorded a very well measured almost vertical air shower with an energy estimated of about $2\times10^{20}$~eV
\cite{agabig}. Later, in 2001 the event of energy of about $2.5\times10^{20}$~eV has been seen in Japan.
However, the world record of the highest UHECR particle energy belongs to the event measured by precursor of the HiRes, the Fly's Eye experiment in 1991. The value to beat is, since then, still $3.2\times10^{20}$~eV \cite{fe3e20}.

\begin{figure}[th]
\centerline{
\includegraphics[width=7cm]{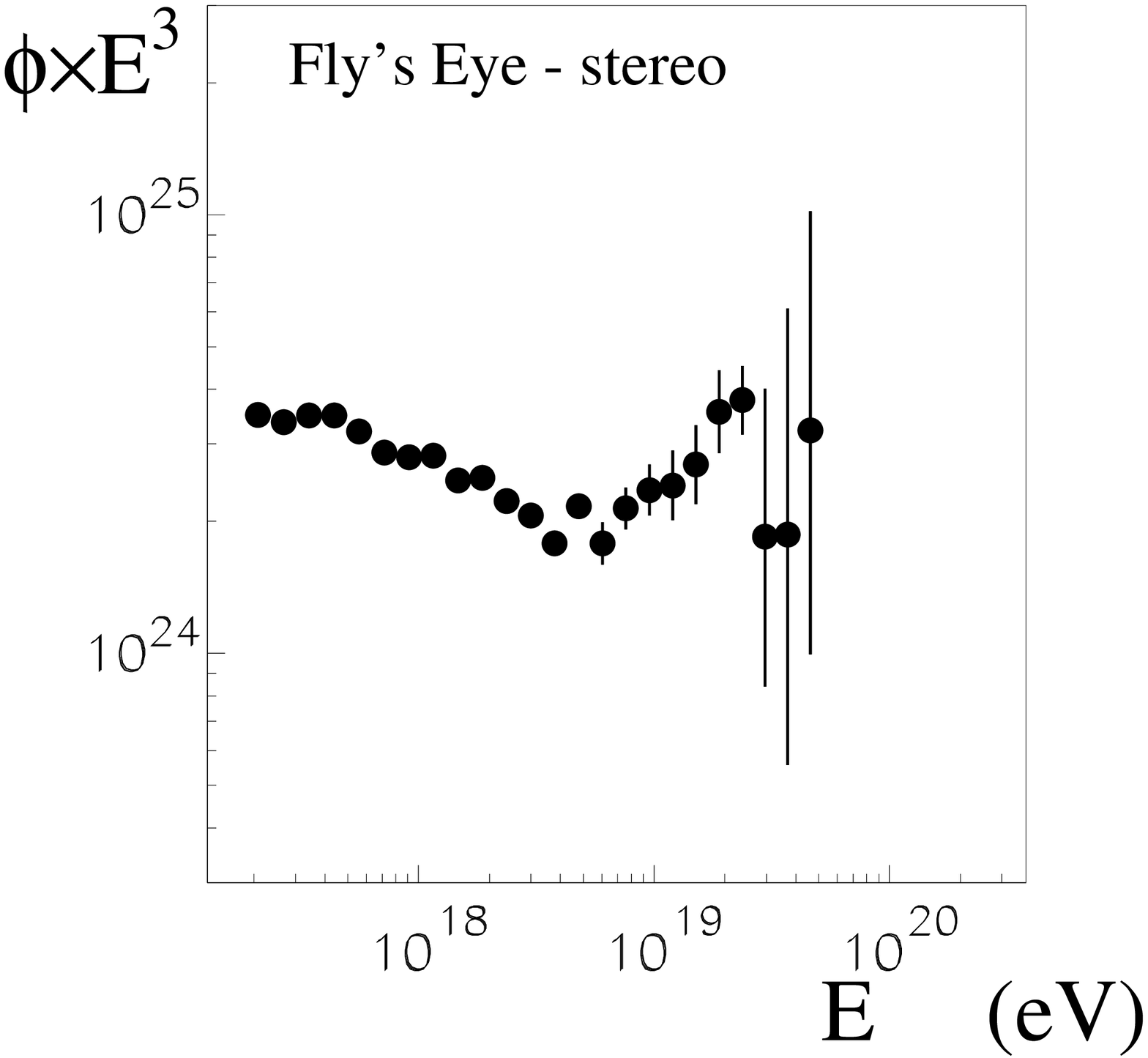}
\includegraphics[width=7cm]{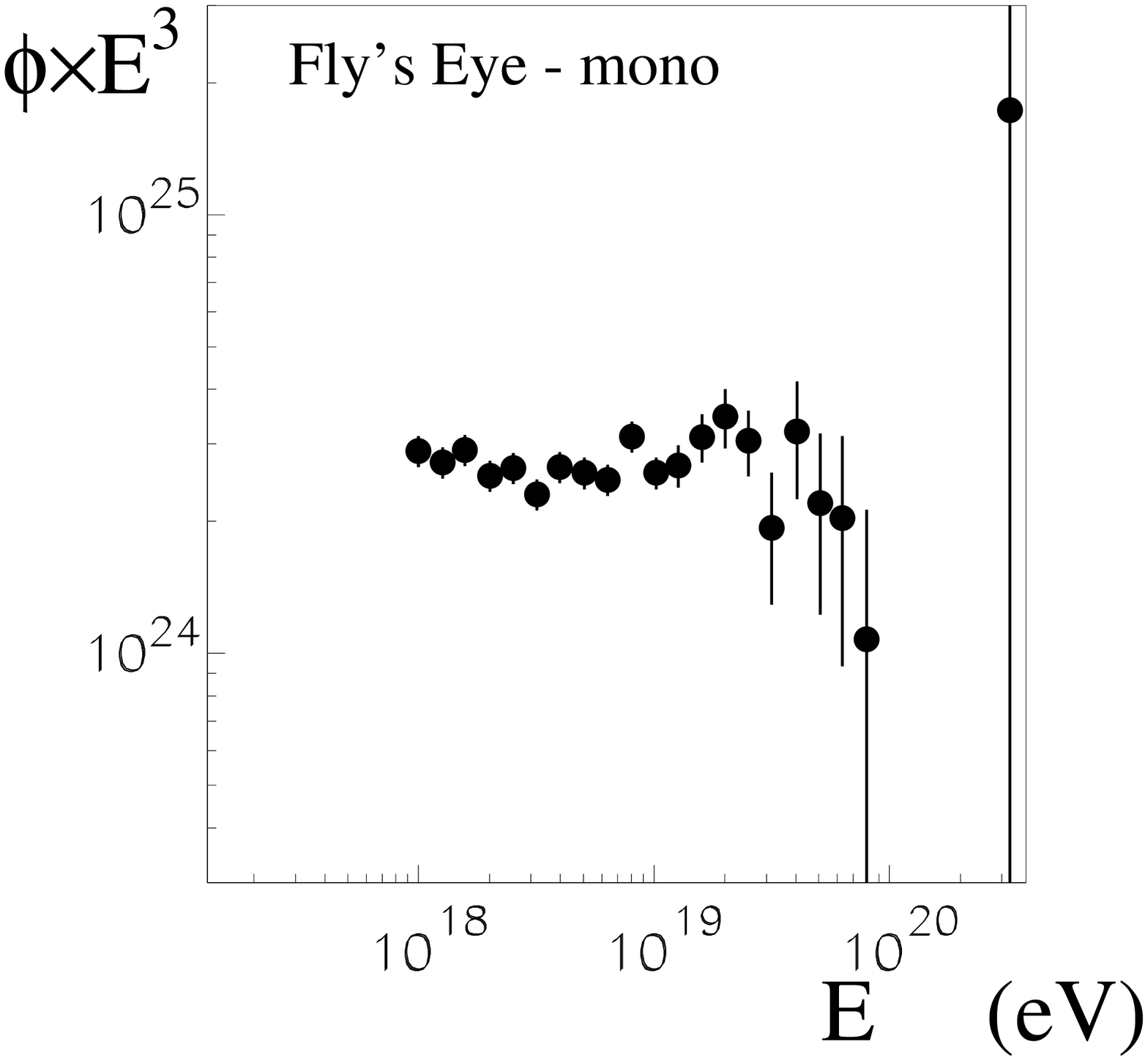}}
\caption{Fly's Eye stereo and mono spectra. 
\label{fe}}
\end{figure}

The Fly's Eye detector begin operation in 1981 first as single Eye monitoring the scintillation flashes 
produced by extremely big cascade of charged particles created by UHECR in the atmosphere. 
In 1985 the second Eye (twice smaller) joint the first completed the apparatus. 
It increased significantly the
geometrical reconstruction procedures accuracy thus the particle energy determination.
The statistics collected by the First Fly's Eye detector, the monocular data set, is much larger than the 
stereo data set. 
%The spectrum determined by the larges sample of events 
%naturally extends much to the high energies. 
Both spectra are shown in Fig.~\ref{fe} 
\cite{fe}. The mentioned single event of energy of $3.2\times10^{20}$~eV is seen as a separated single point in the Fig.~\ref{fe}b.

\begin{figure}[bh]
\centerline{
\includegraphics[width=7cm]{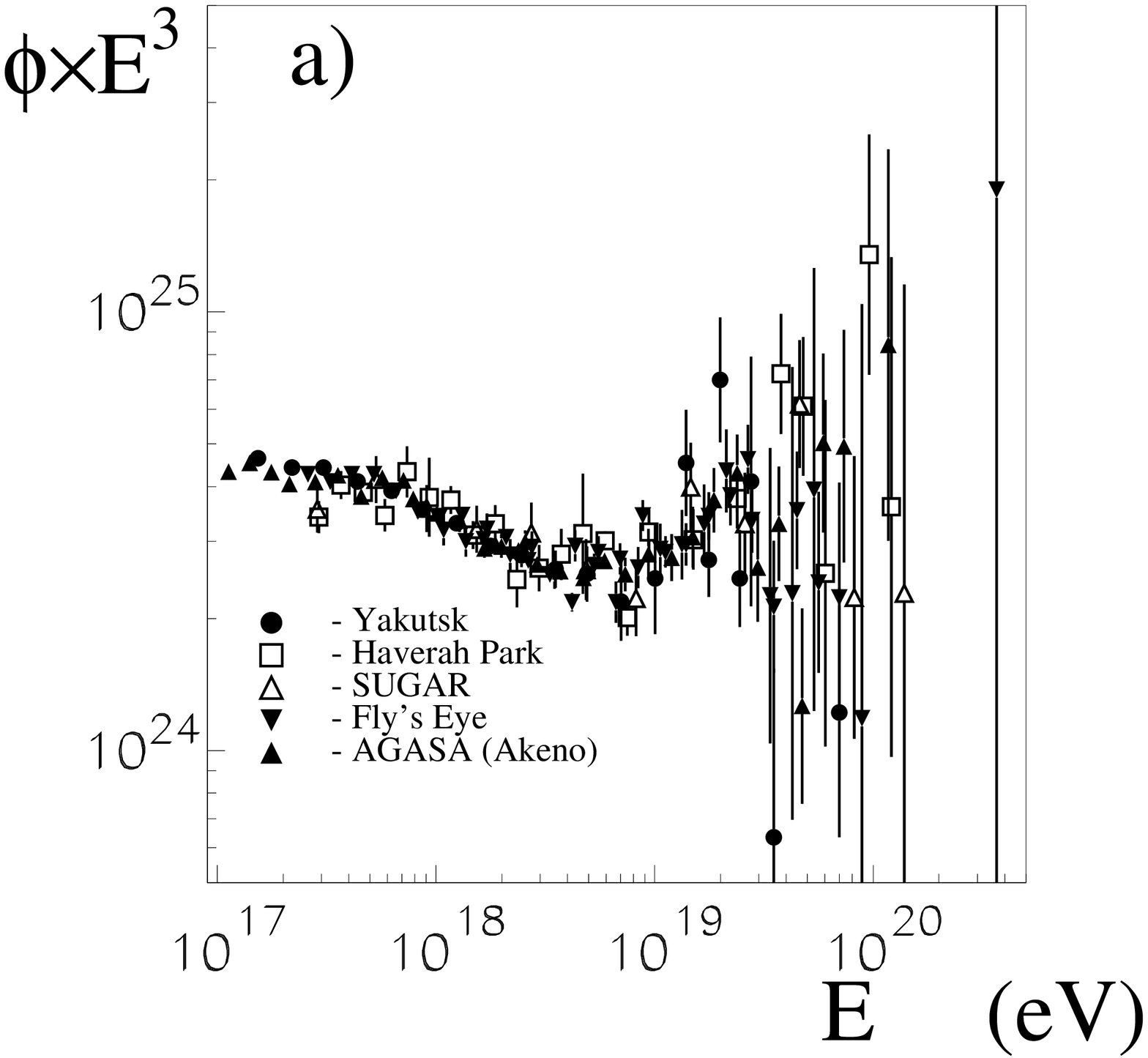}
\includegraphics[width=7cm]{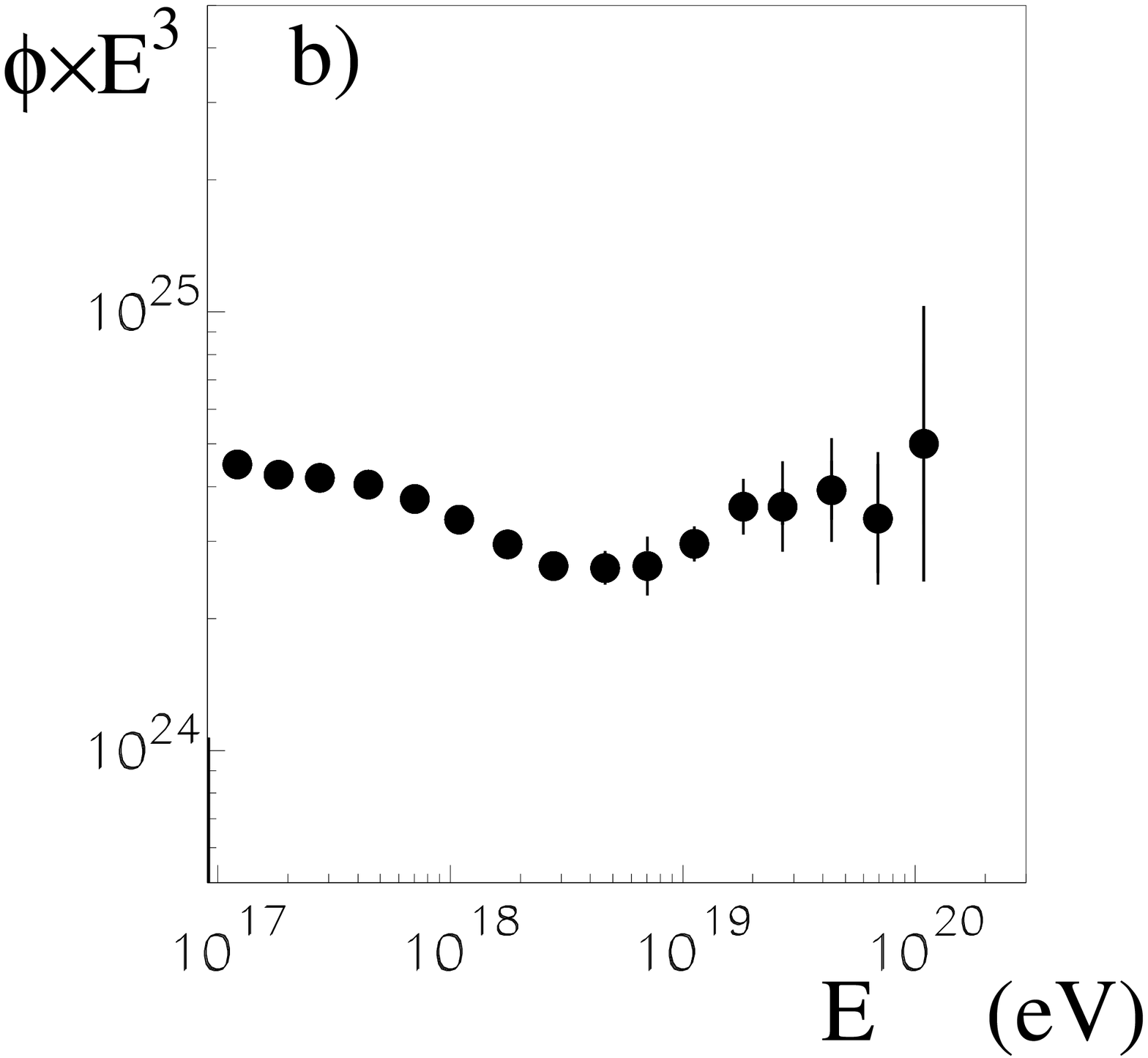}}
\caption{ a) Spectra from Haverah Park, SUGAR, AGASA, Akeno, Fly's Eye and Yakutsk adjusted to 
the same shape at the dip below 10${19}$~eV (the procedure described in \cite{szab}),
and b) the sum of all listed data - the {\it prior}.
\label{prior}}
\end{figure}

There is well known that the spectra from different experiments are displaced, as well in absolute 
energy calibration, as in the normalization of the total measured UHECR flux. In Ref.\cite{szab} the procedure
was developed to shift (and adjust according to different individual energy resolution)
results of each group using the assumption about the universality of the dip structure observed 
in all data sets below 10$^{19}$~eV. In our opinion this feature is related to the change from the Galactic to 
the Extragalactic flux component. There are also other opinions, but no matter what is the source of the dip,
it can be used as an energy and total flux re-calibration. 
When it is done, the spread of the points is 
Gaussian and the average value makes sense, thus the observed spread can be used to determine the error 
of the average. 

The averaging procedure was applied to all data with two exceptions. The one is the Pierre Auger Observatory (PAO) recent spectrum \cite{Matthiae:2008en} published
after the result of HiRes comes out - if not the final `5 sigma' statement in 
Ref.\cite{hires}, than as in the form of an announcement of the GZK cut-off discovery in Ref.\cite{hires-mex}.
The PAO spectrum will be discussed in Sec.\ref{impr} .
%shown in Fig.~\ref{pao} 
The second are HiRes spectra
which are the subject of this paper, and which will be discussed in details in the next section. 
In the Fig.~\ref{prior} the result of the summary is shown. 

The 'world average' presented in Fig.~\ref{prior}b forms exactly the {\it prior} needed in Bayesian reasoning
treatment of probability. The world record event from Fly's Eye mono and the 
Volcano Ranch first UHECR showers are not included in the shown {\it prior}.

\section{The Likelihood \label{lakli}}

\begin{figure}[th]
\centerline{
\includegraphics[width=7cm]{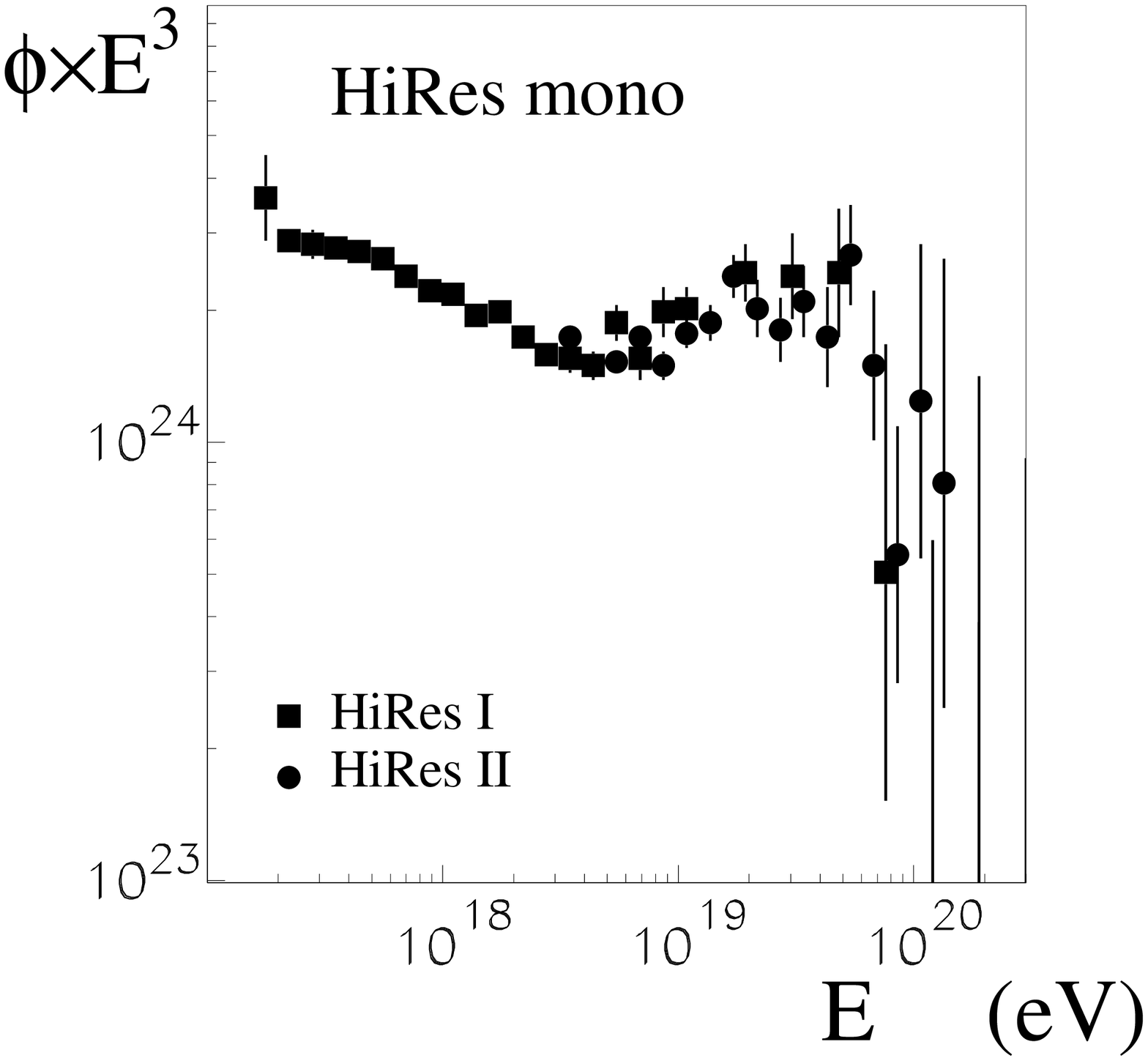}
\includegraphics[width=7cm]{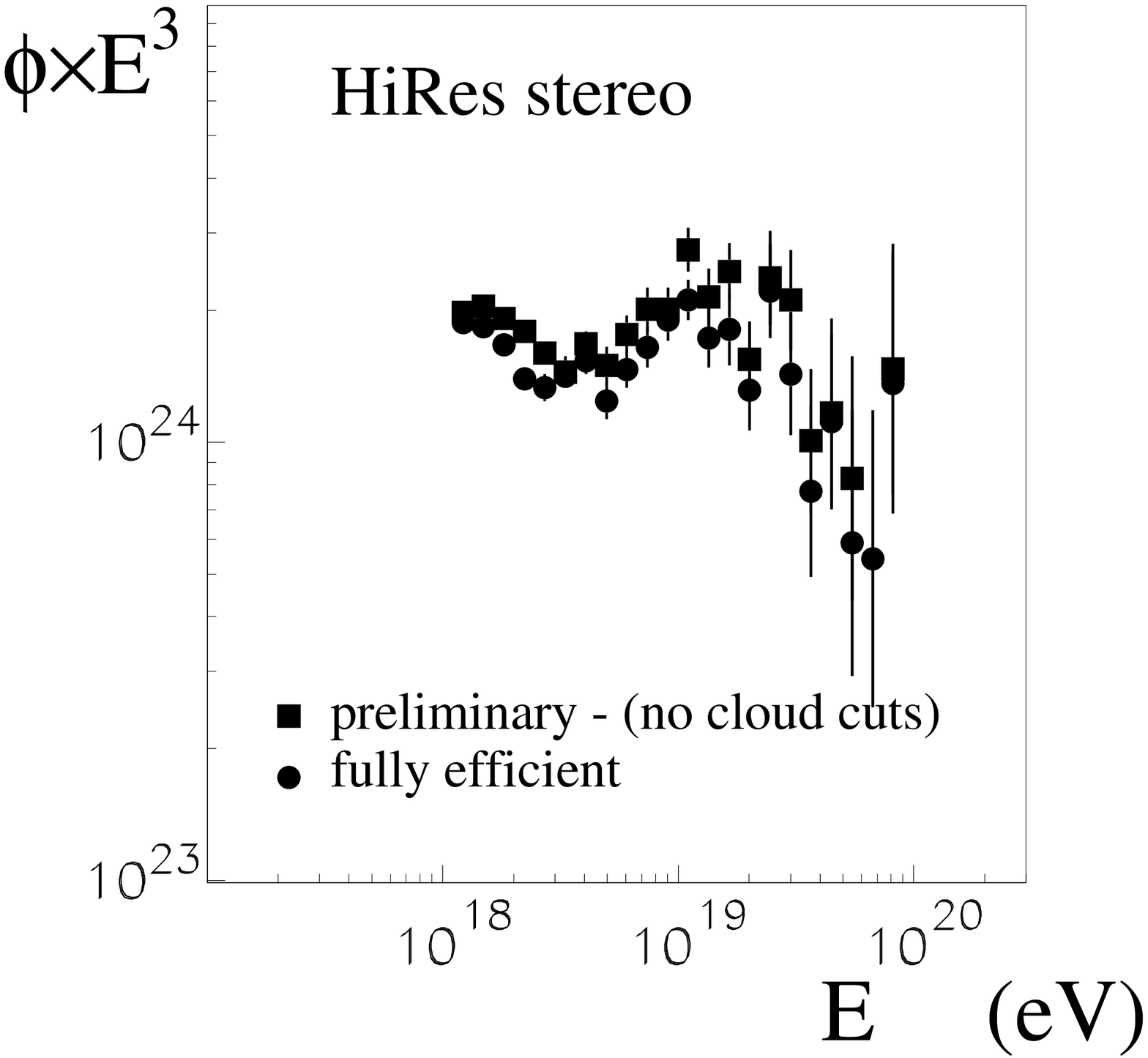}}
\caption{HiRes mono and stereo spectra \label{hiresspe}}
\end{figure}

The factor next to the {\it prior} in a Bayes formula is the likelihood. It describe the
increase of our knowledge related to the new measurement, in our case the recent HiRes experiment spectrum
\cite{hires} shown in Fig.~\ref{hiresspe}. Left panel shows results of two 'mono' spectra of events registered separately by the Eye I and II. In the right the more accurate, but statistically poorer the 'stereo' spectrum obtained from events seen by both Eyes is shown. 

Lets first discuss the '5 sigma' statement as it is published in Ref.~\cite{Abbasi:2007sv}. It is based on the Fig.~\ref{hiresspe}a.

It was obtained comparing two numbers: 43.2 expected and 13 events observed. The comparison was done using the Poisson distribution with the average, expected value, 43.2,
and indeed the probability of observing 13 or less events is on the level of $7\times 10^{-8}$. 

The value of 43.2 expected events is obtained extrapolating the spectrum with the same index as
adjusted to the data between the energies of about $10^{18.5}$~eV and the point of abrupt GZK break announced at $10^{19.75}$~eV. The accuracy of this index reported
is equal to $0.03$ \cite{Abbasi:2007sv}. The change of the index from 2.81 to 2.81+0.03 
increases the likelihood $P(E|H)$ about twice, which doesn't seem to be much. 

The change of the estimated position of the GZK break in the spectrum ($19.75 \pm 0.04$ \cite{Abbasi:2007sv})
from 19.75 to 19.79 decrease expectations from 43.2 to about 36.5 decreasing likelihood significantly. To preserve its level of $7\times 10^{-8}$ the number of observed events should change from 13 to 9 (1/3 of events observed above GZK cut-off has to have energies not more than 10\% higher than the cut-off energy 
$5.6\times 10^{19}$). It is still possible. Such error in the energy determination is within the accuracy of 
the method of energy determination by the experiment, so it is hard to estimate with the actual measured 
sample of 13 events. 

All this details can change some probabilities we are still close to the level of '5 sigma'. The data 
collected by HiRes mono experiments produce likelihood of about $10^{-7}$.

\section{The improvement \label{impr}}

According to the conventional wisdom and, formally, to the Bayes formula, any new measurement should improve 
our knowledge. 
\begin{figure}[bh]
\centerline{
\includegraphics[width=6cm]{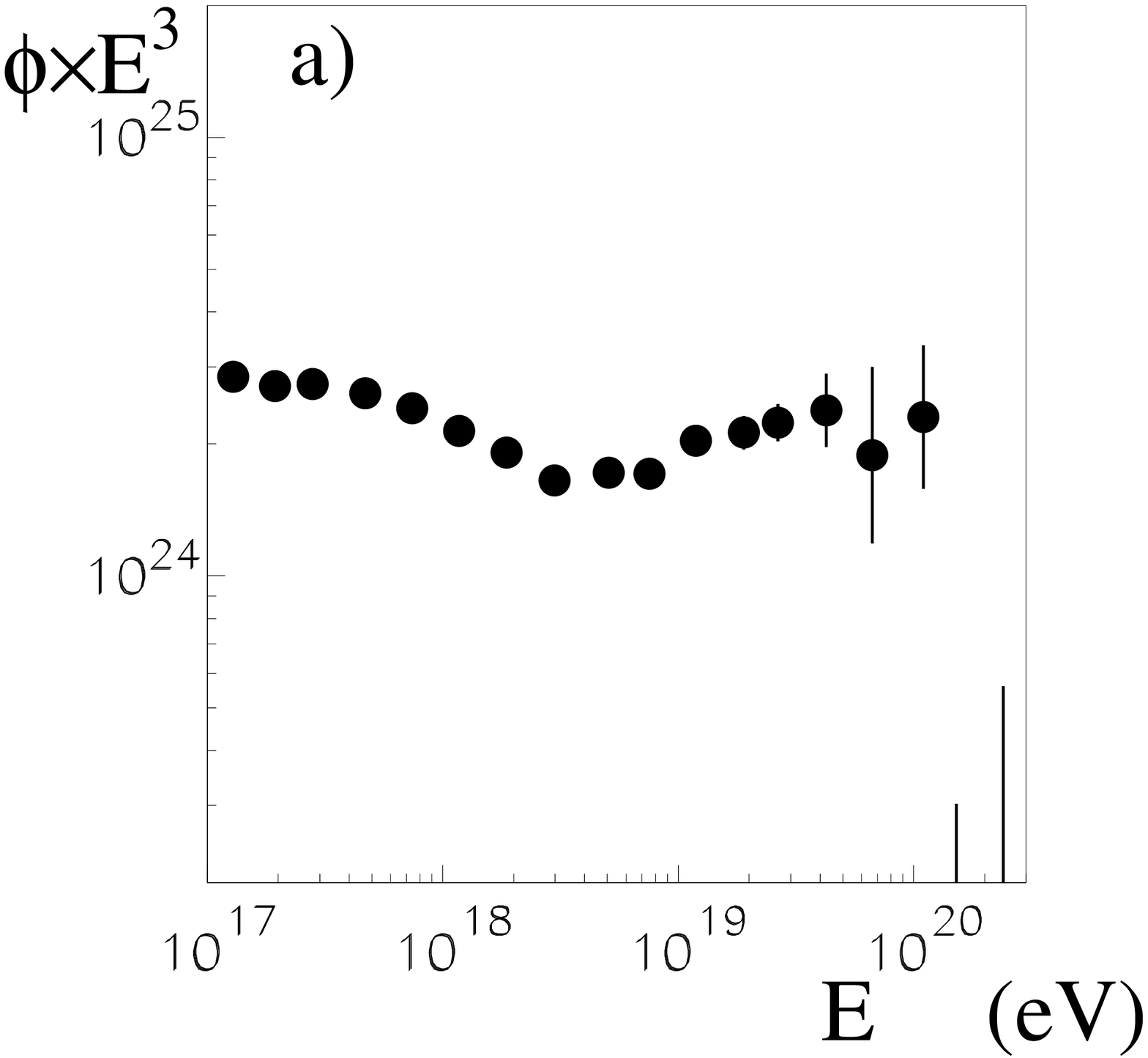}
\includegraphics[width=6cm]{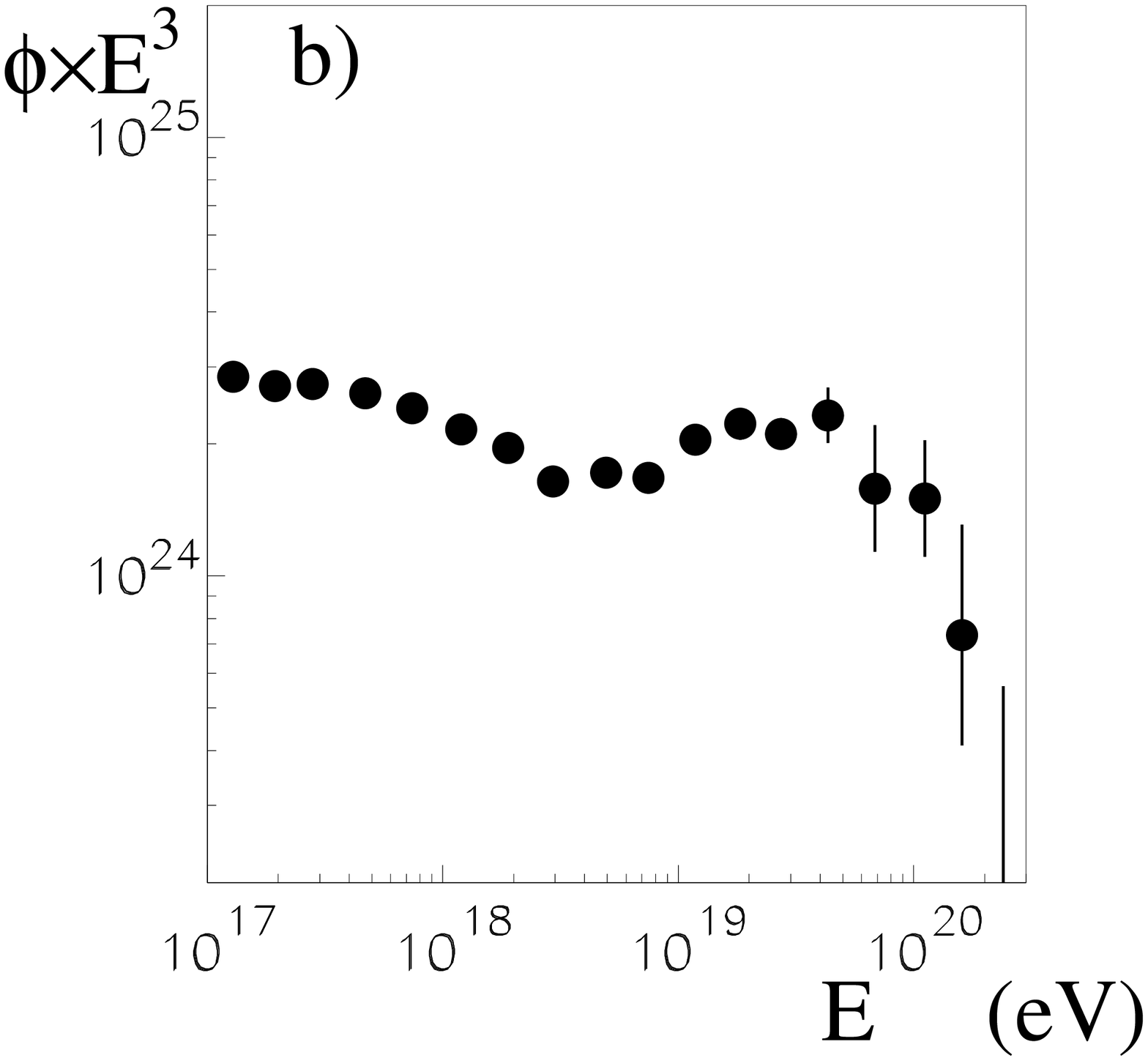}
\includegraphics[width=6cm]{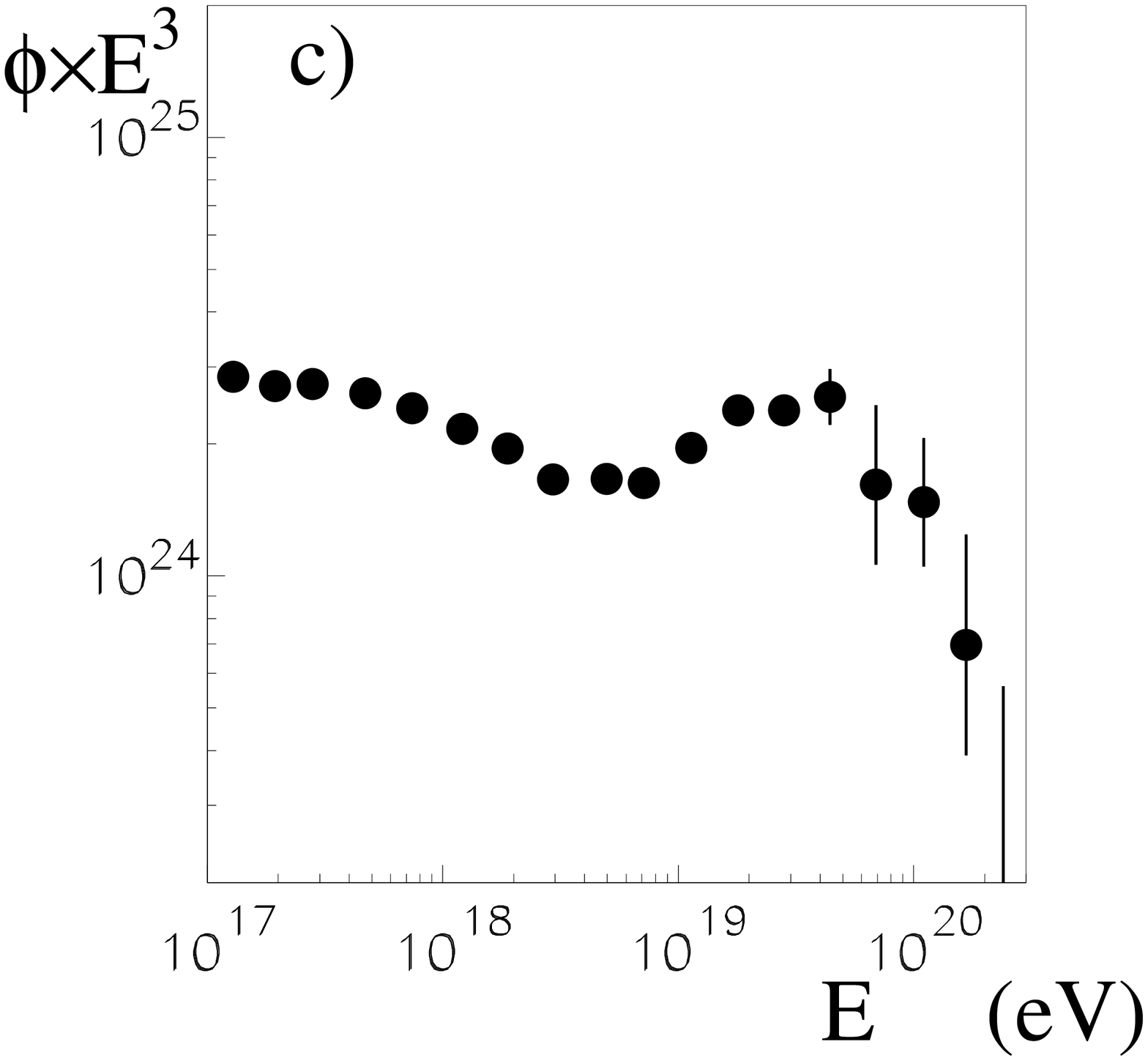}}
\caption{UHECR spectra  a) combined HiRes mono and the {\it prior} from \ref{prior}b, b) HiRes mono and stereo and the {\it prior},
and c)  HiRes and Auger and the {\it prior} -- the 'world average spectrum'.
\label{po}}
\end{figure}

The procedure of averaging UHECR spectra used to obtain the prior (Fig.~\ref{prior}b) can be used of course 
also for all experiments including HiRes mono and stereo, as well as PAO.

The result is presented in Fig.\ref{po}. 

The Fig.\ref{po}a shows the result of combining the HiRes mono spectra \cite{Abbasi:2007sv}
and the {\it prior}. The prior is moved to match the
dip structure where it is seen well in the HiRes data. 

It has been said, that the HiRes mono data itself gives the `5 sigma' confidence. 
From the Bayesian point of view this statement represents 
the situation when nothing else about UHECR flux is known. As 
it has been shown in Sec.\ref{sprior} we already know quite a lot and the real 
spectrum. Accordingly, the GZK significance estimation {\it after} the
HiRes (mono) measurement, can be estimated from the {\it posterior} spectrum shown in Fig.\ref{po}a. 
To get some numbers we follow the procedure used by HiRes 
group in Ref.~\cite{Abbasi:2007sv}. First we get the index of the
UHECR particle spectrum above the ankle where is seems to be stable and where there is no signs of any
cut-off. We used the same energy interval 
$18.5 \le \log_{10} E \le 19.75$
as it was used in Ref.~\cite{Abbasi:2007sv}. 
The value of the index found is 2.87$\pm$0.08.

Then we estimated the probability that, when there is no GZK cut-off, 
the measured flux at the very tail of the spectrum
is at least as low as our {\it improved} UHECR flux.
%We used the last four points in Fig.\ref{po}a. They extend to about $E=3\times  10^{20}$~eV. 
The `absence of the GZK cut-off' means that the UHECR flux 
above $\log_{10} E = 19.75$ continues the
trend found at the ankle below this energy. 
This probability could be estimated with the help of the $\chi^2$ statistics. 
The obtained value is $\chi^2/NDF = 8.7/4$
what gives the chance probabilities of about $3 \times 10^{-2}$. 

Introducing the HiRes stereo data gives the {\it posterior} flux shown in Fig.\ref{po}b. 
Estimated index below the
$\log_{10} E =19.75$ remains unchanged. The complete HiRes (mono+stereo) gives
 $\chi^2/NDF = 7.4/4$, and the chance probability even bigger: $6 \times 10^{-2}$. 

We can say than if the HiRes data are 
combined with the {\it prior} that the existence of the GZK cut-off is observed with the significance 
below 2$\sigma$ level.

\begin{figure}[th]
\centerline{
\includegraphics[width=7cm]{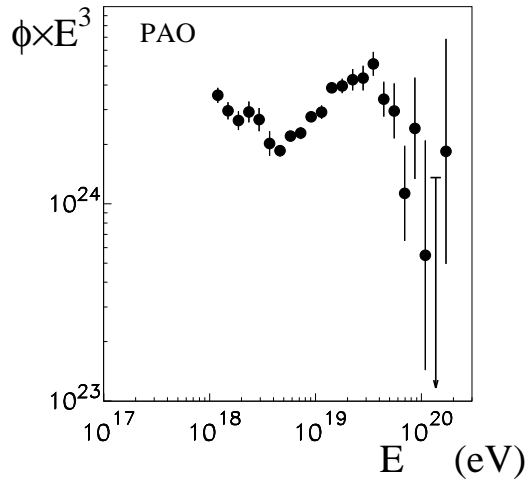}}
\caption{Pierre Auger Observatory UHECR spectrum \cite{Matthiae:2008en}. 
\label{pao}}
\end{figure}

There is, mentioned above, one more data set on UHECR spectrum available for some time. This is Pierre Auger Observatory result
\cite{Matthiae:2008en} shown in Fig.~\ref{pao}. 
The PAO spectrum exhibits also the cut-off 
in general accordance with HiRes result, in spite of its energy calibration 
and absolute flux normalization.
The enhancement of the probability in favour of the GZK picture is expected.

It is not very substantial as it can be seen in Fig.~\ref{po}c, 
where the combination of all measured spectra is shown. 
The index before the GZK threshold energy is slightly changed to 2.86$\pm 0.13$, and
the chance probability changes to 
$1.4 \times 10^{-2}$ and the significance 
level of the GZK cut-off discovery, expressed in sigmas, exceeded 2$\sigma$.

It should be mentioned here the existence of the very high energy events 
not included in the present analysis.  
The first ever super-GZK Volcano Ranch event discussed in Sec.~\ref{sprior}, 
the mentioned `world record' Fly's Eye event, 
and the one registered by PAO just on the edge (but slightly outside) of 
the working part of the array, 
additionally diminish the GZK cut-off existence probability.

\subsection{Observation of the evolutionary effect \label {zapo}}

It is interesting to note the evolution of the UHECR flux results measured by different experiments. 
It is in general nothing extraordinary. It is well known, e.g., in the lasting case of gravitational constant.
Another example is given in the first one 
of Particle Data Group history plots: the neutron life-time case. Error boxes 
of results measured before 1970 are far outside the nowadays accepted value \cite{groom}.

The UHECR flux Haverah Park measurement history 
is given in Fig.\ref{hp789} while the Yakutsk in Fig.\ref{yak}. Both exhibit similar effect. 
The initially published spectra do not follow the GZK hypothesis, just opposite, they 
continue gradually.

\begin{figure}[th]
\centerline{
\includegraphics[width=7cm]{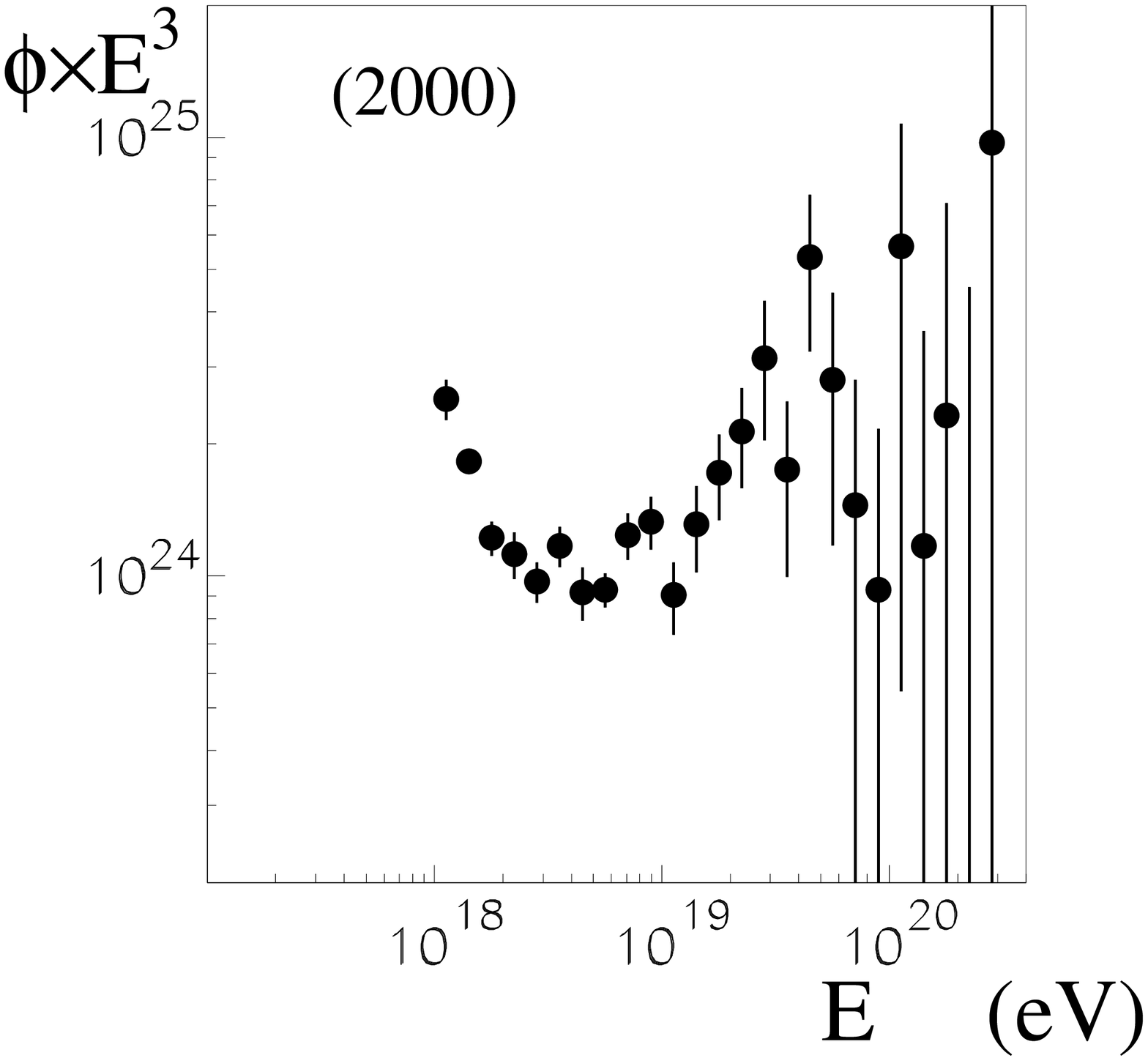}
\includegraphics[width=7cm]{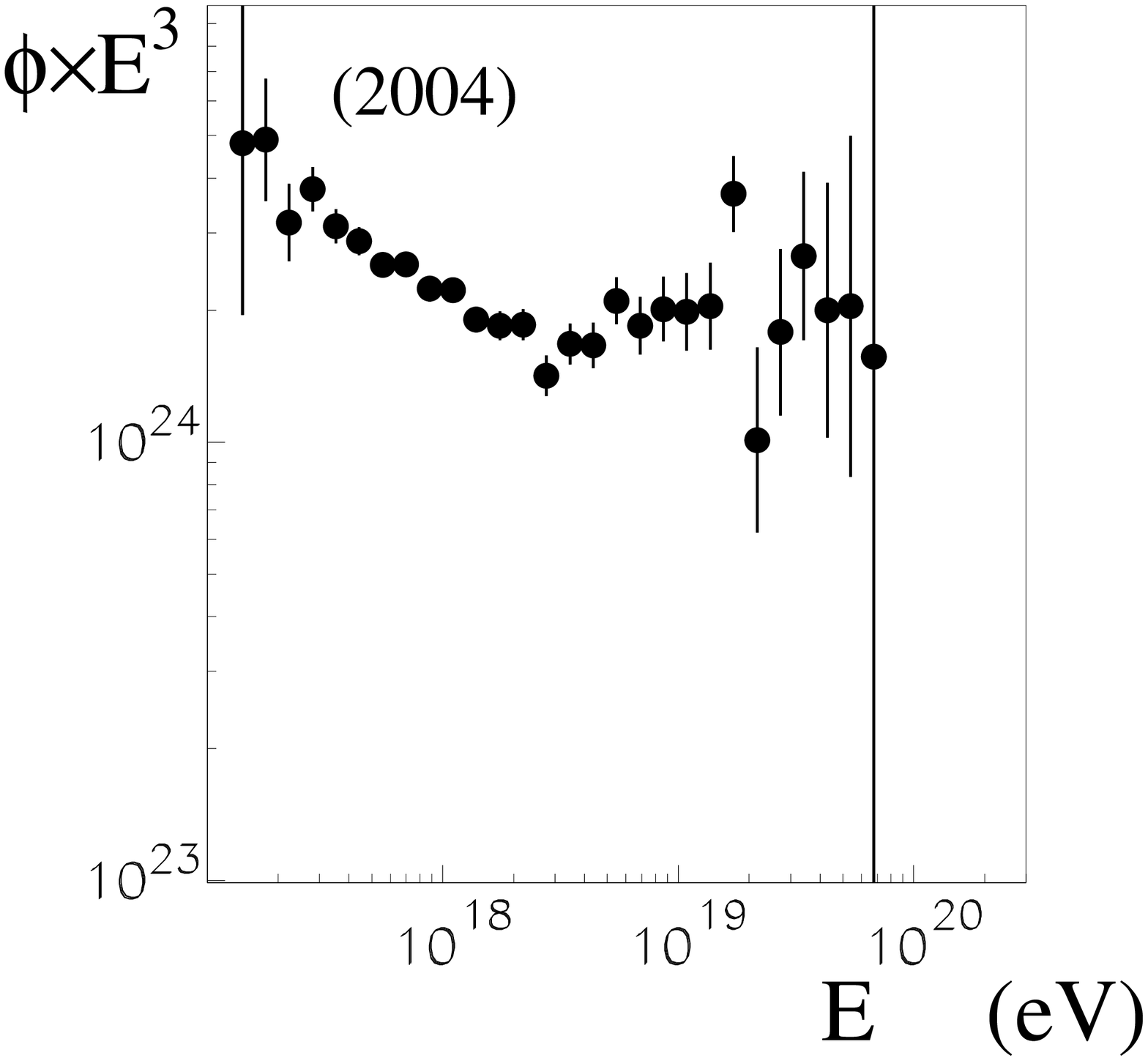}}
\caption{(left) HiRes spectrum collected in 1997-1999 \cite{abuza}  and (right) the spectrum published in \cite{zech}.
\label{abuzech}}
\end{figure}

The spectra of HiRes telescopes shown in
Fig.\ref{hiresspe} are its recent version showing `5 sigma' deficit of super-GZK events. 
At the end of previous century HiRes spectrum looked quite different. In Fig.\ref{abuzech}a 
the spectrum of HiRes I (BigH) 
detector is shown as it was published in Ref.\cite{abuza}. There are 13 
events observed between May 1997 to June 1999 with energy exceeding 
$6\times 10^{19}$~eV and 7 with energy greater $10^{20}$~eV!
All details of these super-GZK events are published in Ref.~\cite{abuza}.

HiRes data few years ago (but in XXI century) looked also slightly different. Fig.\ref{abuzech}b presents
the HiRes II spectrum obtained from data collected between
December 1999 and September 2001 \cite{zech}. 

The probability of the GZK cut-off hypothesis is increasing in last years, not only 
according to the new measurements but also confirmed by the re-analysis of the old data.

The `evolution' of the UHECR spectra is sustained by another interesting fact. 
The knowledge gathered by the experimental group not so long ago, even in '80, vanishes recently very fast, 
in a sense. This possibility was mentioned in Sec.~\ref{proba}
The contemporary new and big experiments, much bigger than the old ones, and hundreds of 
people working there 
draw their conclusions like there was nothing before them. But the super-GZK events seen some time ago
remain the super-GZK still, even if one doesn't like them. 

The similar conclusion can be found in the recent paper analysing the UHECR data
by Glushkov and Pravdin \cite{glusz}.

\section{Summary and Conclusions}

The rational concluding the facts presented in the previous Section we have to take into account 
non-physical factor, which for frequentalists 
sounds like insult, and it is neutral for Bayesians. The factor of {\it believe}.

The believe in GZK (thus believe in extragalactic protons) acts as additional {\it prior} and
influence the reasoning driving to the conclusions which are (or, in general, could be) wrong. 
Wrong, from 'frequentalist', physical point of view, which they (the frequentalists)
believe, should be free of any believes.

If one would bet 10000000 to 1 for the GZK cut-off existence, thus the pure proton flux in the UHECR domain
than his additional, `non-physical' believe in GZK, 
is on the level of 10000 to 1. 
This is simply Bayesian conclusion of the calculations presented above. 
The `5 sigma' effect combined with the prior (and additionally with the PAO result) turns 
out to be only little above `2 sigma'.

As a conclusion we have shown that the GZK cut-off if exist, 
has the overall significance of about `2 sigma' far less than `5 sigma'. Thus the claim that the
Greisen and Zatzepin and Kuzmin has been established experimentally is a little premature.

%The similar point of view can be found in the recent paper of Glushkov and Pravdin \cite{glusz}.

\end{document}